\documentclass[prb,preprintnumbers,amsmath,amssymb,floatfix]{revtex4}

\usepackage{graphicx}
\usepackage{subfigure}
\usepackage{dcolumn}

\makeatletter

\begin{document}
\baselineskip 16pt

\title{Function Photonic Crystals }
\author{Xiang-Yao Wu$^{a}$ \footnote{E-mail: wuxy2066@163.com},
Bai-Jun Zhang$^{a}$, Jing-Hai Yang$^{a}$, Xiao-Jing Liu$^{a}$\\
Nuo Ba$^{a}$, Yi-Heng Wu$^{a}$, Qing-Cai Wang$^{a}$ and Guang-Huai
Wang$^{a}$}

 \affiliation{a.Institute of Physics, Jilin Normal
University, Siping 136000 }

\begin{abstract}
In the paper, we present a new kind of function photonic crystals,
which refractive index is a function of space position. Unlike
conventional PCs, which structure grow from two materials, A and
B, with different dielectric constants $\varepsilon_{A}$ and
$\varepsilon_{B}$. Based on Fermat principle, we give the motion
equations of light in one-dimensional, two-dimensional and
three-dimensional function photonic crystals. For one-dimensional
function photonic crystals, we investigate the dispersion
relation, band gap structure and transmissivity, and compare them
with conventional photonic crystals. By choosing various
refractive index distribution function $n(z)$, we can obtain more
wider or more narrower band gap structure than conventional
photonic crystals.

\end{abstract}
\maketitle

\maketitle {\bf 1. Introduction} \vskip 8pt

Photonic crystals (PCs), proposed by Yablonovitch and John,
represent a novel class of optical materials which allow to
control the flow of electromagnetic radiation or to modify
light-matter interaction [1, 2]. These artificial structures are
characterized by one, two or three-dimensional arrangements of
dielectric material which lead to the formation of an energy band
structure for electromagnetic waves propagating in them. One of
the most attractive features of photonic crystals is associated
with the fact that PCs may exhibit frequency ranges over which
ordinary linear propagation is forbidden, irrespective of
direction. These photonic band gaps (PBGs) lend themselves to
numerous diversified applications (in linear, nonlinear and
quantum optics). For instance, PBG structures with line defects
can be used for guiding light. Similarly, as it has been predicted
and confirmed experimentally, photonic crystals allow to modify
spontaneous emission rate due to the modification of density of
quantum states. In particular, it is well known that the density
of states grows at the edge of the photonic band gap of the PCs.
This allows us to predict a higher optical gain, but on the other
hand a higher level of noise in light generated in PC-lasers
operated at a frequency near the band gap.

Photonic crystals are usually viewed as an optical analog of
semiconductors that modify the properties of light similarly to a
microscopic atomic lattice that creates a semiconductor band gap
for electrons [3]. It is therefore believed that by replacing
relatively slow electrons with photons as the carriers of
information, the speed and bandwidth of advanced communication
systems will be dramatically increased, thus revolutionizing the
telecommunication industry. To employ the high-technology
potential of photonic crystals, it is crucially important to
achieve a dynamical tunability of their band gap [4]. This idea
can be realized by changing the light intensity in the so-called
nonlinear photonic crystals, having a periodic modulation of the
nonlinear refractive index [5]. Exploration of nonlinear
properties of photonic band-gap (PBG) materials could be exploited
new applications of photonic crystals to devise all-optical signal
processing and switching, which indicates an effective way to
generate tunable band-gap structures by operating entirely with
light.

During the past few years, there has been a great deal of interest
in studying propagation of waves inside periodic structures. These
systems are composites made of inhomogeneous distribution of some
material periodically embedded in other with different physical
properties. Phononic crystals (PCs)[6, 7] are one of the examples
of these systems. PCs are the extension of the so-called Photonic
crystals[8] when elastic and acoustic waves propagate in periodic
structures made of materials with different elastic properties.
When one of these elastic materials is a fluid medium, then PCs
are called Sonic Crystals (SC)[9, 10]. For these artificial
materials, both theoretical and experimental results have shown
several interesting physical properties [11]. In the
homogenization limit [12], it is possible to design acoustic
metamaterials that can be used to build refractive devices [13].
In the range of wavelengths similar to the periodicity of the PCs,
multiple scattering process inside the PC leads to the phenomenon
of so called Band Gaps (BG),which are required for filtering sound
[10], trapping sound in defects [14, 15] and for acoustic wave
guiding [16].

In the present work we present a new kind of function photonic
crystals, which refractive index is a function of the space
position. Unlike conventional PCs, which grow from two materials,
A and B, with different dielectric constants $\varepsilon_{A}$ and
$\varepsilon_{B}$: a periodic layered medium $...A/B/A/B...$ in
case of one-dimensional photonic crystals and periodic arrays of
cylinders and spheres of the material A embedded in a dielectric
matrix B, in case of two-dimensional and three-dimensional
photonic crystals, respectively. Function PCs may extend the
concept of PCs, leading likely to some new applications. To
exemplify the idea of function PCs, we present theoretical
calculations of the function photonic crystals band structures and
transmissivity. Our results indicate that the function photonic
crystals behaves more width or more narrow band gap structure than
the conventional photonic crystals.

\vskip 8pt

{\bf 2. The light motion equation in function photonic crystals}
\vskip 8pt

For the function photonic crystals, the crystals refractive index
is a periodic function of the space position, which can be written
as $n(z)$, $n(x, z)$ $n(x, y, z)$ corresponding to
one-dimensional, two-dimensional and three-dimensional function
photonic crystals. In the following, we shall deduce the light
motion equations for the one-dimensional, two-dimensional and
three-dimensional function photonic crystals. Firstly, we give the
light motion equation in one-dimensional function photonic
crystals, and two-dimensional motion space, i.e., the refractive
index is $n=n(z)$, meanwhile motion path is on $xz$ plane. The
incident light wave strikes plane interface $A$ point, the curves
$AB$ and $BC$ are the path of incident and reflected light
respectively, and they are shown in FIG. 1.
\begin{figure}[tbp]
\includegraphics[width=9 cm]{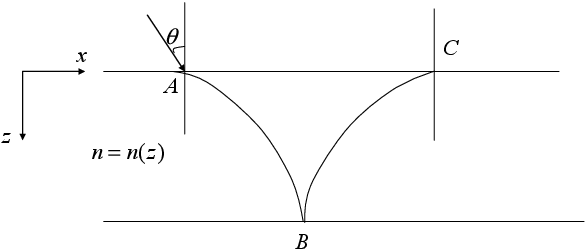}
\caption{The motion path of light in one-dimensional function
photonic crystals and two-dimensional motion space} \label{Fig1}
\end{figure}

The light motion equation can be obtained by Fermat principle, and
it is
\begin{eqnarray}
\delta\int^{B}_{A}n(z) ds=0.
\end{eqnarray}
In the two-dimensional transmission space, the line element $ds$
is
\begin{eqnarray}
ds=\sqrt{(dx)^{2}+(dz)^{2}}=\sqrt{1+\dot{z}^{2}}dx,
\end{eqnarray}
where $\dot{z}=\frac{dz}{dx}$, then Eq. (1) becomes
\begin{eqnarray}
\delta\int^{B}_{A}n(z)\sqrt{1+(\dot{z})^{2}}dx=0.
\end{eqnarray}
The Eq. (3) change into
\begin{eqnarray}
\int^{B}_{A}(\frac{\partial(n(z)\sqrt{1+\dot{z}^{2}})}{\partial
z}\delta z+\frac{\partial(n(z)\sqrt{1+\dot{z}^{2}})}{\partial
\dot{z}}\delta\dot{z})dx=0,
\end{eqnarray}
i.e.,
\begin{eqnarray}
 \nonumber\\&&\int^{B}_{A}\frac{d n(z)}{d
z}\sqrt{1+\dot{z}^{2}}\delta
zdx+\int^{B}_{A}n(z)\dot{z}(1+\dot{z}^{2})^{-\frac{1}{2}}d\delta z
\nonumber\\&=&\int^{B}_{A}\frac{d n(z)}{d
z}\sqrt{1+\dot{z}^{2}}\delta
zdx+n(z)\dot{z}(1+\dot{z}^{2})^{-\frac{1}{2}}\delta z|^{B}_{A}
-\int^{B}_{A}d(n(z)\dot{z}(1+\dot{z}^{2})^{-\frac{1}{2}})\delta
z\nonumber\\&=&0.
\end{eqnarray}
The two end points $A$ and $B$, their variation is zero, i.e.,
$\delta z (A)=\delta z (B)=0$, and the Eq. (5) is \\
\begin{eqnarray}
\int^{B}_{A} (\frac{d n(z)}{d z}\sqrt{1+\dot{z}^{2}} -\frac{d
n(z)}{d z}\dot{z}^{2}(1+\dot{z}^{2})^{-\frac{1}{2}}
-n(z)\frac{\ddot{z}\sqrt{1+\dot{z}^{2}}
-\dot{z}^{2}\ddot{z}(1+\dot{z}^{2})^{-\frac{1}{2}}}{1+\dot{z}^{2}})
\delta zdx=0.
\end{eqnarray}
For arbitrary variation $\delta z$, there is
\begin{eqnarray}
\frac{dn(z)}{dz}\sqrt{1+\dot{z}^{2}} -\frac{d n(z)}{d
z}\dot{z}^{2}(1+\dot{z}^{2})^{-\frac{1}{2}}
-n(z)\frac{\ddot{z}\sqrt{1+\dot{z}^{2}}
-\dot{z}^{2}\ddot{z}(1+\dot{z}^{2})^{-\frac{1}{2}}}{1+\dot{z}^{2}}
=0,
\end{eqnarray}
simplify Eq. (7), we have
\begin{eqnarray}
\frac{dn(z)}{dz} -n(z)\frac{\ddot{z}}{1+\dot{z}^{2}}=0.
\end{eqnarray}\vskip 15pt
The Eq. (8) is light motion equation in one-dimensional function
photonic crystals and two-dimensional motion space. Similarly, we
can attain light motion equation in one-dimensional function
photonic crystals and three-dimensional motion space. It is
\begin{eqnarray}
\frac{dn(z)}{dz}\dot{y}(1+\dot{y}^{2}+\dot{z}^{2})-n(z)(\dot{y}\ddot{z}-\dot{z}\ddot{y})=0.
\end{eqnarray}
For the two-dimensional function photonic crystals, light in
two-dimensional motion space, the light motion equation is
\begin{eqnarray}
\frac{\partial n(x,z)}{\partial
z}-n(x,z)\frac{\ddot{z}}{1+\dot{z}^2}-\frac{\partial
n(x,z)}{\partial x}\dot{z}=0,
\end{eqnarray}
and the light in three-dimensional motion space, the motion
equation is
\begin{eqnarray}
\frac{\partial n(x,z)}{\partial z}\dot{y}(1+\dot{y}^{2}+
\dot{z}^{2})-n(x,z)(\dot{y}\ddot{z}-\dot{z}\ddot{y})=0.
\end{eqnarray}
For the three-dimensional function photonic crystals, light in
two-dimensional motion space, the light motion equation is
\begin{eqnarray}
\frac{\partial n(x,y_{0},z)}{\partial z}-\frac{\partial
n(x,y_{0},z)}{\partial
x}\dot{z}-n(x,y_{0},z)\frac{\ddot{z}}{1+\dot{z}^{2}}=0,
\end{eqnarray}
where $y_{0}$ is constant, and the light transmits in
three-dimensional motion space, the light motion equation is
\begin{eqnarray}
\nonumber\\&&(1+\dot{y}^{2}+\dot{z}^{2})(\frac{\partial n(x, y,
z)}{\partial y}\dot{z}-\frac{\partial n(x,y,z)}{\partial
z}\dot{y}) -n(x,y,z)(\ddot{y}\dot{z}-\ddot{z}\dot{y})=0.
\end{eqnarray}

{\bf 3. The transfer matrix of one-dimensional function photonic
crystals} \vskip 8pt

In this section, we should calculate the transfer matrix of
one-dimensional function photonic crystals and two-dimensional
motion space. In fact, it is the reflection and refraction of
light at a plane surface of two media with different dielectric
properties. The dynamic properties of the electric field and
magnetic field are contained in the boundary conditions: normal
components of $D$ and $B$ are continuous; tangential components of
$E$ and $H$ are continuous. We consider the electric field
perpendicular to the plane of incidence, and the coordinate system
and symbols as shown in FIG. 2.

\begin{figure}[tbp]
\includegraphics[width=8 cm]{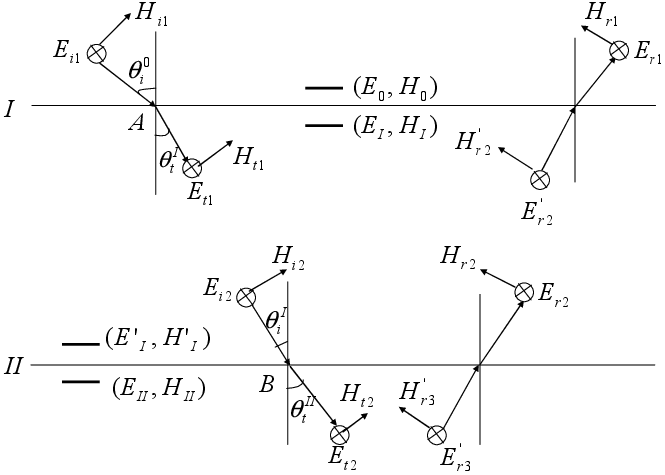}
\caption{The light transmission figure in arbitrary middle medium}
\label{Fig1}
\end{figure}
On the two sides of interface I, the tangential components of
electric field $E$ and magnetic field $H$ are continuous, there
are
\begin{eqnarray}
\left \{ \begin{array}{ll}
 E_{0}=E_{I}=E_{t1}+E'_{r2}\\
H_{0}=H_{I}=H_{t1}\cos\theta_{t}^{I}-H'_{r2}\cos\theta_{t}^{I},
\end{array}
\right.
\end{eqnarray}
On the two sides of interface II, the tangential components of
electric field $E$ and magnetic field $H$ are continuous, and give
\begin{eqnarray}
\left \{ \begin{array}{ll}
 E_{II}=E'_{I}=E_{i2}+E_{r2}\\
H_{II}=H'_{I}=H_{i2}\cos\theta_{i}^{I}-H_{r2}\cos\theta_{i}^{I},
\end{array}
\right.
\end{eqnarray}
the electric field ${E_{t1}}$ is
\begin{eqnarray}
E_{t1}=E_{t10}{e^{i(k_{x}x_{A}+k_{z}z)}|_{z=0}}=E_{t10}e^{i\frac{\omega}{c}n(0)\sin\theta_{t}^{I}x_{A}},
\end{eqnarray}
and the electric field ${E_{i2}}$ is
\begin{eqnarray}
\nonumber\\&&E_{i2}=E_{t10}{e^{i(k'_{x}x_{B}+k'_{z} z)}|_{z=b}}
=E_{t10}e^{i\frac{\omega}{c}n(b)(\sin\theta_{i}^{I}x_{B}+\cos\theta_{i}^{I}
b)}.
\end{eqnarray}
Where $x_{A}$ and $x_{B}$ are $x$ component coordinates
corresponding to $A$
and $B$ points.\\
Now, we calculate the incident angle $\theta_{i}^{I}$. From Eq.
(8), it is straightforward to derive
\begin{eqnarray}
\frac{d n(z)}{n(z)}=\frac{\dot{z}d\dot{z}}{1+\dot{z}^{2}},
\end{eqnarray}
then integrate the two sides of Eq. (18)
\begin{eqnarray}
\int^{n(b)}_{n(0)}\frac{dn(z)}{n(z)}=\int^{k_{b}}_{k_{0}}\frac{\dot{z}d\dot{z}}{1+\dot{z}^{2}},
\end{eqnarray}
to get
\begin{eqnarray}
(\frac{n(b)}{n(0)})^{2}=\frac{1+k_{b}^2}{1+k_{0}^2},
\end{eqnarray}
where
\begin{eqnarray}
k_{b}={\frac{dz}{dx}}|_{z=b}=\cot\theta_{i}^{I},
\end{eqnarray}
\begin{eqnarray}
k_{0}={\frac{dz}{dx}}|_{z=0}=\cot\theta_{t}^{I},
\end{eqnarray}
and
\begin{eqnarray}
\theta_{t}^{I}=\arcsin(\frac{n_{0}}{n(0)}\sin\theta_{i}^{0}),
\end{eqnarray}
where $n_0$ is air refractive index, and $n(0)=n(z)|_{z=0}$.\\ By
substituting Eqs. (21), (22) and (23) into (20), we attain
\begin{eqnarray}
\frac{1+\cot^{2}\theta_{i}^{I}}{1+\cot^{2}\theta_{t}^{I}}=\frac{n^{2}(b)}{n^{2}(0)}.
\end{eqnarray}
From Eq. (24), we can find when $n(0)=n(b)$, there is
\begin{eqnarray}
\theta_{t}^{I}=\theta_{i}^{I}.
\end{eqnarray}
Integrate the two sides of Eq. (18), we can obtain the coordinate
component $x_{B}$
\begin{eqnarray}
\int^{n(z)}_{n(0)}\frac{dn(z)}{n(z)}=\int^{k_{z}}_{k_{0}}\frac{\dot{z}d\dot{z}}{1+\dot{z}^{2}},
\end{eqnarray}
to get
\begin{eqnarray}
(\frac{n(z)}{n(0)})^{2}=\frac{1+k_{z}^{2}}{1+k_{0}^{2}},
\end{eqnarray}
since $k_{z}> 0$, there is
\begin{eqnarray}
k_{z}=\frac{dz}{dx}=\sqrt{(1+k_{0}^{2})(\frac{n(z)}{n(0)})^{2}-1},
\end{eqnarray}
i.e.,
\begin{eqnarray}
dx=\frac{dz}{\sqrt{(1+k_{0}^{2})(\frac{n(z)}{n(0)})^{2}-1}}.
\end{eqnarray}
Obviously, $n(z)>n(0)\sin\theta^{I}_{t}$. The coordinate $x_{B}$
can be obtained as
\begin{eqnarray}
x_{B}=x_{A}+\int^{b}_{0}\frac{dz}{\sqrt{(1+k_{0}^{2})(\frac{n(z)}{n(0)})^{2}-1}}.
\end{eqnarray}
By substituting Eq. (30) into (17), there is
\begin{eqnarray}
\nonumber\\&&E_{i2}=E_{t10}\exp[{i\frac{\omega}{c}n(b)(\sin\theta_{i}^{I}
\int^{b}_{0}\frac{dz}{\sqrt{(1+k_{0}^{2})(\frac{n(z)}{n(0)})^{2}-1}}+\sin\theta_{i}^{I}x_{A}+\cos\theta_{i}^{I}b)}],
\end{eqnarray}
With substituting Eq. (16) into (31), there is
\begin{eqnarray}
E_{i2}&=&E_{t1}e^{i{\delta}_{b}},
\end{eqnarray}
where
\begin{eqnarray}
\delta_{b}=\frac{\omega}{c}n(b)(\cos\theta_{i}^{I}b+\sin\theta_{i}^{I}
\int^{b}_{0}\frac{dz}{\sqrt{(1+k_{0}^{2})(\frac{n(z)}{n(0)})^{2}-1}}),
\end{eqnarray}
and similarly
\begin{eqnarray}
E'_{r2}=E_{r2}e^{i\delta_{b}}.
\end{eqnarray}
Substituting Eqs. (32) and (34) into (14) and (15), and using
$H=\sqrt{\frac{\varepsilon_{0}}{\mu_{0}}}nE$, there are
\begin{eqnarray}
\left \{ \begin{array}{ll}
E_{I}=E_{t1}+E_{r2}e^{i\delta_{b}}\\
H_{I}=\sqrt{\frac{\varepsilon_{0}}{\mu_{0}}}n(0)E_{t1}\cos\theta_{t}^{I}
-\sqrt{\frac{\varepsilon_{0}}{\mu_{0}}}n(0)E_{r2}\cos\theta_{t}^{I}e^{i\delta_{b}},
 \end{array}
   \right.
\end{eqnarray}
and
\begin{eqnarray}
\left \{ \begin{array}{ll}
E_{II}=E_{t1}e^{i\delta_{b}}+E_{r2}\\
H_{II}=\sqrt{\frac{\varepsilon_{0}}{\mu_{0}}}n(b)E_{t1}e^{i\delta_{b}}\cos\theta_{i}^{I}
-\sqrt{\frac{\varepsilon_{0}}{\mu_{0}}}n(b)E_{r2}\cos\theta_{i}^{I}.
\end{array}
   \right.
\end{eqnarray}
From Eq. (35) and (36), we can obtain
\begin{eqnarray}
 \left \{ \begin{array}{ll}
E_{I}=\frac{1}{\sqrt{\frac{\varepsilon_{0}}{\mu_{0}}}n(b)\cos\theta_{i}^{I}}
(\sqrt{\frac{\varepsilon_{0}}{\mu_{0}}}n(b)\cos\theta_{i}^{I}\cos\delta_{b}E_{II}
-i \sin\delta_{b}H_{II})\\
H_{I}=\frac{n(0)\cos\theta^{I}_{t1}}{n(b)\cos\theta^{I}_{i2}}
 (-i\sqrt{\frac{\varepsilon_{0}}{\mu_{0}}}n(b)\cos\theta_{i}^{I}
\sin\delta_{b}E_{II}+\cos\delta_{b}H_{II}),
  \end{array}
   \right.
\end{eqnarray}
or
\begin{eqnarray}
\left(%
\begin{array}{c}
  E_{I} \\
  H_{I} \\
\end{array}%
\right)=\left(%
\begin{array}{cc}
 \cos\delta_{b} & -\frac{i\sin\delta_{b}}{\sqrt{\frac{\varepsilon_{0}}{\mu_{0}}}n(b)
  \cos\theta_{i}^{I}}\\
  -i\sqrt{\frac{\varepsilon_{0}}{\mu_{0}}}n(0)
  \cos\theta_{t}^{I}\sin\delta_{b}
   & \frac{n(0)\cos\theta_{t}^{I}}{n(b)\cos\theta_{i}^{I}}\cos\delta_{b}\\
\end{array}%
\right)\left(%
\begin{array}{c}
  E_{II} \\
  H_{II} \\
\end{array}%
\right),
\end{eqnarray}
define $M$ matrix
\begin{eqnarray}
\left(%
\begin{array}{c}
  E_{I} \\
   H_{I} \\
\end{array}%
\right)=M\left(%
\begin{array}{c}
  E_{II}  \\
  H_{II}  \\
\end{array}%
\right)
\end{eqnarray}
where
\begin{eqnarray}
M=\left(%
\begin{array}{cc}
 \cos\delta_{b} & -\frac{i}{\sqrt{\frac{\varepsilon_{0}}{\mu_{0}}}n(b)
  \cos\theta_{i}^{I}}\sin\delta_{b} \\
 -i\sqrt{\frac{\varepsilon_{0}}{\mu_{0}}}n(0)
  \cos\theta_{t}^{I}\sin\delta_{b}
 & \frac{n(0)\cos\theta_{t}^{I}}{n(b)\cos\theta_{i}^{I}}\cos\delta_{b}.\\
\end{array}%
\right).
\end{eqnarray}
The Eq. (40) is the transfer matrix $M$ of the half period. When
$n(0)=n(b)$, there is
\begin{eqnarray}
\theta_{t}^{I}=\theta_{i}^{I},
\end{eqnarray}
and the $M$ matrix becomes
\begin{eqnarray}
M=\left(%
\begin{array}{cc}
 \cos\delta_{b} & -\frac{i}{\eta_{b}}\sin\delta_{b} \\
 -i\eta_{b}\sin\delta_{b}
 & \cos\delta_{b}\\
\end{array}%
\right),
\end{eqnarray}
where
$\eta_{b}=\sqrt{\frac{\varepsilon_{0}}{\mu_{0}}}n(0)\cos\theta_{i}^{I}$.
In the following, we should study the one-dimensional function
Photonic crystals at $n(0)=n(b)$.

\vskip 8pt

{\bf 4. The structure of one-dimensional function photonic
crystals} \vskip 8pt

In section 3, we attain the $M$ matrix of the half period. We know
that the conventional photonic crystals is constituted by two
different refractive index medium, and the refractive indexes are
not continuous on the interface of the two medium. We could devise
the one-dimensional function photonic crystals structure as
follows: in the first half period, the medium refractive index is
$n_{1}(z)$, and in the second half period, the medium refractive
index is $n_{2}(z)$, corresponding thickness $b$ and $a$,
respectively. Their refractive indexes satisfy condition
$n_{1}(b)\neq n_{2}(0)$, as shown in FIG. 3. We should
discuss two kinds of incidence cases: \\
(1) The light vertical incidence

For the vertical incidence, the initial incidence angle
$\theta_{i}^{0}=0$ and refraction angle $\theta_{t}^{I}=0$. From
Eq. (41), there is
\begin{eqnarray}
\theta_{i}^{I}=0,
\end{eqnarray}
then
\begin{eqnarray}
\theta_{t}^{II}=0, \hspace{0.3in} \theta_{i}^{II}=0,
\hspace{0.3in} \theta_{t}^{III}=0, \hspace{0.15in}\cdots
\end{eqnarray}
i.e., all incidence angles and refraction angles are zero, and the
$M$ matrix is
\begin{eqnarray}
M=\left(%
\begin{array}{cc}
  \cos\delta_{b} & -\frac{i}{\eta_{b}}\sin\delta_{b} \\
  -i\eta_{b}\sin\delta_{b} & \cos\delta_{b}\\
\end{array}%
\right).
\end{eqnarray}
\begin{figure}[tbp]
\includegraphics[width=8 cm]{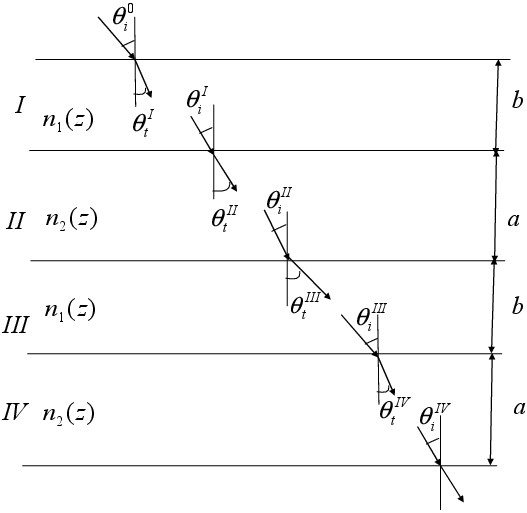}
\caption{The two periods transmission figure of light in function
photonic crystals} \label{Fig1}
\end{figure}
Where $\delta_{b}=\frac{\omega}{c}n(b)b$ and
$\eta_{b}=\sqrt{\frac{\epsilon_{0}}{\mu_{0}}}n(b)$. In the case,
the function photonic crystals becomes conventional common
photonic crystals, which its refractive indexes $n_{a}$ and
$n_{b}$ are constants.\\
(2) The light non-vertical incidence

For light non-vertical incidence, the initial incidence angle
$\theta^{0}_{i}\neq 0$ in the first half period $I$.\\
By refraction law, there is
\begin{eqnarray}
\sin\theta^{I}_{t}=\frac{n_{0}}{n_{1}(0)}\sin\theta^{0}_{i},
\end{eqnarray}
where $n_0$ is air refractive indexes, and
$n_{1}(0)=n_{1}(z)|_{z=0}$.\\
when $n_{1}(0)=n_{1}(b)$, $\theta^{I}_{i}=\theta^{I}_{t}$, we
obtain the $M_{b}$ matrix in the first half period $I$ as
\begin{eqnarray}
M^{I}_{b}=\left(%
\begin{array}{cc}
  \cos\delta^{I}_{b} & -\frac{i}{\eta^{I}_{b}}\sin\delta^{I}_{b} \\
  -i\eta^{I}_{b}\sin\delta^{I}_{b} & \cos\delta^{I}_{b} \\
\end{array}%
\right)
\end{eqnarray}
where
\begin{eqnarray}
\eta^{I}_{b}=\sqrt{\frac{\varepsilon_{0}}{\mu_{0}}}n_{1}(0)\cos\theta^{I}_{t}
\end{eqnarray}
and
\begin{eqnarray}
\delta^{I}_{b}=\frac{\omega}{c}n_{1}(0)[\cos\theta^{I}_{t}\cdot
b+\sin\theta^{I}_{t}\int^{b}_{0}\frac{dz}{\sqrt{(1+k_{0}^{2})(\frac{n_{1}(z)}{n_{1}(0)})^{2}-1}}].
\end{eqnarray}
Where $k_{0}=\cot \theta_{t}^{I}$.\\
In the second half period $II$, by refraction law, there is
\begin{eqnarray}
\sin\theta^{II}_{t}=\frac{n_{1}(b)}{n_{2}(0)}\sin\theta^{I}_{i},
\end{eqnarray}
when $n_{2}(0)=n_{2}(a)$, we have
\begin{eqnarray}
\theta^{II}_{i}=\theta^{II}_{t},
\end{eqnarray}
The $M_{a}$ matrix in the second half period $II$ is obtained
\begin{eqnarray}
M^{I}_{a}=\left(%
\begin{array}{cc}
  \cos\delta^{II}_{a} & -\frac{i}{\eta^{II}_{a}}\sin\delta^{II}_{a} \\
  -i\eta^{II}_{a}\sin\delta^{II}_{a} & \cos\delta^{II}_{a} \\
\end{array}%
\right),
\end{eqnarray}
where
\begin{eqnarray}
\eta^{II}_{a}=\sqrt{\frac{\varepsilon_{0}}{\mu_{0}}}n_{2}(0)\cos\theta^{II}_{t},
\end{eqnarray}
and
\begin{eqnarray}
\delta^{II}_{a}=\frac{\omega}{c}n_{2}(0)[\cos\theta^{II}_{t}\cdot
a+\sin\theta^{II}_{t}\int^{a}_{0}\frac{dz}{\sqrt{(1+k_{I}^{2})(\frac{n_{2}(z)}{n_{2}(0)})^{2}-1}}],
\end{eqnarray}
where $k_{I}=\cot\theta_{t}$. From Eqs. (46) and (50), we have
\begin{eqnarray}
\sin\theta^{II}_{t} =\frac{n_{0}}{n_{2}(0)}\sin\theta^{0}_{i},
\end{eqnarray}
and
\begin{eqnarray}
\cos\theta^{II}_{t}=\sqrt{1-\frac{n_{0}^{2}}{(n_{2}(0))^{2}}\sin^{2}\theta^{0}_{i}}.
\end{eqnarray}
Then the $M$ matrix in the first period is expressed as
\begin{eqnarray}
M^{I}=M^{I}_{b}\cdot M^{I}_{a}
\end{eqnarray}
In the next, we should calculate the $M$ matrix in the second period.\\
By refraction law, there is
\begin{eqnarray}
\sin\theta^{III}_{t}=\frac{n_{2}(b)}{n_{1}(0)}\sin\theta^{II}_{i}
=\frac{n_{0}}{n_{1}(0)}\sin\theta^{0}_{i}=\sin\theta^{I}_{t}
\end{eqnarray}
when $n_{1}(0)=n_{1}(b)$, we have
\begin{eqnarray}
\theta^{III}_{i}=\theta^{III}_{t}=\theta^{I}_{t},
\end{eqnarray}
We can obtain the $M_{b}$ matrix in the second period
\begin{eqnarray}
M^{II}_{b}=\left(%
\begin{array}{cc}
  \cos\delta^{III}_{b} & -\frac{i}{\eta^{III}_{b}}\sin\delta^{III}_{b} \\
  -i\eta^{III}_{b}\sin\delta^{III}_{b} & \cos\delta^{III}_{b} \\
\end{array}%
\right)
\end{eqnarray}
where
\begin{eqnarray}
\eta^{III}_{b}=\sqrt{\frac{\varepsilon_{0}}{\mu_{0}}}n_{1}(0)\cos\theta^{III}_{t}=\eta^{I}_{b},
\end{eqnarray}
\begin{eqnarray}
\delta^{III}_{b}=\frac{\omega}{c}n_{1}(0)[\cos\theta^{III}_{t}\cdot
b+\sin\theta^{III}_{t}\int^{b}_{0}\frac{dz}{\sqrt{(1+k_{II}^{2})(\frac{n_{1}(z)}{n_{1}(0)})^{2}-1}}]=\delta^{1}_{b}.
\end{eqnarray}
From Eqs. (60)-(62), we find
\begin{eqnarray}
M_{b}^{II}=M_{b}^{I}.
\end{eqnarray}
Now, we calculate the $M_{a}$ matrix in
the second period.\\
By refraction law, there is
\begin{eqnarray}
\sin\theta^{IV}_{t}=\frac{n_{1}(b)}{n_{2}(0)}\sin\theta^{III}_{i}
=\frac{n_{0}}{n_{2}(0)}\sin\theta^{0}_{i}=\sin\theta^{II}_{t},
\end{eqnarray}
when $n_{2}(0)=n_{2}(a)$, we have
\begin{eqnarray}
\theta^{IV}_{i}=\theta^{IV}_{t}=\theta^{II}_{t}
\end{eqnarray}
We can obtain the $M_{a}$ matrix in the second half period $IV$ of
the second period
\begin{eqnarray}
M^{II}_{a}=\left(%
\begin{array}{cc}
  \cos\delta^{IV}_{a} & -\frac{i}{\eta^{IV}_{a}} \sin\delta^{IV}_{a} \\
 -i\eta^{IV}_{a}\sin\delta^{IV}_{b} & \cos\delta^{IV}_{b} \\
\end{array}%
\right),
\end{eqnarray}
where
\begin{eqnarray}
\eta^{IV}_{a}=\sqrt{\frac{\varepsilon_{0}}{\mu_{0}}}n_{2}(0)\cos\theta^{IV}_{t}=\eta^{II}_{a},
\end{eqnarray}
and
\begin{eqnarray}
\delta^{IV}_{a}=\frac{\omega}{c}n_{2}(0)
[\cos\theta^{IV}_{t}a+\sin\theta^{IV}_{t}\int^{a}_{0}\cot(\ln\frac{n_{2}(z)}{n_{2}(0)}+\arctan(\cot\theta^{IV}_{t}))dz]
=\delta^{II}_{a}.
\end{eqnarray}
From Eqs. (66)-(68), we find
\begin{eqnarray}
M_{a}^{II}=M_{a}^{I}.
\end{eqnarray}
By calculation, we find that all $M_{a}$ matrixes are equal, and
all $M_{b}$ matrixes are also equal in different period, they are
\begin{eqnarray}
M_{a}=\left(%
\begin{array}{cc}
  \cos\delta_{a} & -\frac{i}{\eta_{a}} \sin\delta_{a} \\
 -i\eta_{a}\sin\delta_{a} & \cos\delta_{a} \\
\end{array}%
\right),
\end{eqnarray}
and
\begin{eqnarray}
M_{b}=\left(%
\begin{array}{cc}
  \cos\delta_{b} & -\frac{i}{\eta_{b}} \sin\delta_{b} \\
 -i\eta_{b}\sin\delta_{b} & \cos\delta_{b} \\
\end{array}%
\right),
\end{eqnarray}
where
\begin{eqnarray}
\eta_{a}=\sqrt{\frac{\varepsilon_{0}}{\mu_{0}}}n_{2}(0)\sqrt{1-\frac{n^{2}_{0}}{n^{2}_{2}(0)}\sin^{2}\theta^{0}_{i}},
\end{eqnarray}
\begin{eqnarray}
\eta_{b}=\sqrt{\frac{\varepsilon_{0}}{\mu_{0}}}n_{1}(0)\sqrt{1-\frac{n^{2}_{0}}{n^{2}_{1}(0)}\sin^{2}\theta^{0}_{i}},
\end{eqnarray}
\begin{eqnarray}
\delta_{a}&=&\frac{\omega}{c}n_{2}(0) [\cos\theta^{II}_{t}\cdot
a+\sin\theta^{II}_{t}\int^{a}_{0}\frac{dz}{\sqrt{(1+k_{I}^{2})(\frac{n_{2}(z)}{n_{2}(0)})^{2}-1}}],
\end{eqnarray}
and
\begin{eqnarray}
\delta_{b}=\frac{\omega}{c}n_{1}(0) [\cos\theta^{I}_{t}\cdot
b+\sin\theta^{I}_{t}\int^{b}_{0}\frac{dz}{\sqrt{(1+k_{0}^{2})(\frac{n_{1}(z)}{n_{1}(0)})^{2}-1}}],
\end{eqnarray}
where $k_{0}=\cot \theta_{t}^{I}$ and $k_{I}=\cot
\theta_{t}^{II}$.\\
Finally, we obtain the $M$ matrix for every period
\begin{eqnarray}
M=M_{b}M_{a}=\left(%
\begin{array}{cc}
   \cos\delta_{b} & -\frac{i}{\eta_{b}} \sin\delta_{b} \\
  -i\eta_{b}\sin\delta_{b} & \cos\delta_{b}\\
\end{array}%
\right)\left(%
\begin{array}{cc}
   \cos\delta_{a} & -\frac{i}{\eta_{a}} \sin\delta_{a} \\
  -i\eta_{a}\sin\delta_{a} & \cos\delta_{a}\\
\end{array}%
\right).
\end{eqnarray}
For the N-th period, the field vector of up and down $E_{N}$,
$H_{N}$ and $E_{N+1}$, $E_{N+1}$ are satisfied with characteristic
equation
\begin{eqnarray}
\left(%
\begin{array}{c}
  E_{N} \\
  H_{N} \\
\end{array}%
\right)=M_{N}
\left(%
\begin{array}{c}
  E_{N+1} \\
  H_{N+1} \\
\end{array}%
\right).
\end{eqnarray}
From Eq. (77), we can further obtain the characteristic equation
of the $N$ periods photonic crystals, it is
\begin{eqnarray}
\left(%
\begin{array}{c}
  E_{1} \\
  H_{1} \\
\end{array}%
\right)&=&M_{1}M_{2}\cdot\cdot\cdot M_{N}
\left(%
\begin{array}{c}
  E_{N+1} \\
  H_{N+1} \\
\end{array}%
\right) =M_{b}M_{a}M_{b}M_{a}\cdot\cdot\cdot M_{b}M_{a}\left(%
\begin{array}{c}
  E_{N+1} \\
  H_{N+1} \\
\end{array}%
\right)
\nonumber\\&=&M\left(%
\begin{array}{c}
  E_{N+1} \\
  H_{N+1} \\
\end{array}%
\right)=\left(%
\begin{array}{c c}
  A &  B \\
 C &  D \\
\end{array}%
\right)
 \left(%
\begin{array}{c}
  E_{N+1} \\
  H_{N+1} \\
\end{array}%
\right).
\end{eqnarray}

{\bf 5. The dispersion relation, band gap structure and
transmissivity} \vskip 8pt

With the transfer matrix $M$ (Eq. (76)), we can study the
dispersion relation and band gap structure of the function
photonic
crystals.\\
From Eqs. (70) and (71), there is
\begin{eqnarray}
\left(%
\begin{array}{c}
  E_{N} \\
 H_{N}\\
\end{array}%
\right)=M_{b}M_{a}\left(%
\begin{array}{c}
  E_{N+1} \\
  H_{N+1} \\
\end{array}%
\right).
\end{eqnarray}
By Bloch law, we have
\begin{eqnarray}
\left(%
\begin{array}{c}
  E_{N} \\
 H_{N}\\
\end{array}%
\right)=e^{-ikd}\left(%
\begin{array}{c}
  E_{N+1} \\
  H_{N+1} \\
\end{array}%
\right),
\end{eqnarray}
where $d=b+a$. With Eqs. (79) and (80), there is
\begin{eqnarray}
\left(%
\begin{array}{c}
  E_{N} \\
  H_{N} \\
\end{array}%
\right)=M_{b}M_{a}\left(%
\begin{array}{c}
  E_{N+1} \\
  H_{N+1} \\
\end{array}%
\right)=e^{-ikd}\left(%
\begin{array}{c}
  E_{N+1} \\
  H_{N+1} \\
\end{array}%
\right),
\end{eqnarray}
The non-zero solution condition of Eq. (81) is
\begin{eqnarray}
det(M_{b}M_{a}-e^{-ikd})=0,
\end{eqnarray}
i.e.,
\begin{eqnarray}
\nonumber\\&&(\cos\delta_{b}\cos\delta_{a}-\frac{\eta_{a}}{\eta_{b}}\sin\delta_{b}\sin\delta_{a}-e^{-ikd})
(\cos\delta_{b}\cos\delta_{a}-\frac{\eta_{b}}{\eta_{a}}\sin\delta_{b}\sin\delta_{a}-e^{-ikd})
\nonumber\\&&+(-\frac{i}{\eta_{a}}\cos\delta_{b}\sin\delta_{a}-\frac{i}{\eta_{b}}\sin\delta_{b}\cos\delta_{a})
(-i\eta_{b}\sin\delta_{b}\cos\delta_{a}-i\eta_{a}\cos\delta_{b}\sin\delta_{a})=0,
\end{eqnarray}
we resolve the dispersion relation
\begin{eqnarray}
\cos
kd=\cos\delta_{b}\cos\delta_{a}-\frac{1}{2}(\frac{\eta_{b}}{\eta_{a}}+\frac{\eta_{a}}{\eta_{b}})\sin\delta_{b}\sin\delta_{a},
\end{eqnarray}
From Eq. (84), we can study the photonic dispersion relation and
band gap structure, and we can obtain the transmission coefficient
$t$ from Eq. (78)
\begin{eqnarray}
t=\frac{E_{tN+1}}{E_{i1}}=\frac{2\eta_{0}}{A\eta_{0}+B\eta_{0}\eta_{N+1}+C+D\eta_{N+1}},
\end{eqnarray}
and transmissivity $T$
\begin{eqnarray}
T=t\cdot t^{*}
\end{eqnarray}

\newpage

\begin{figure}[tbp]
\includegraphics[width=10 cm]{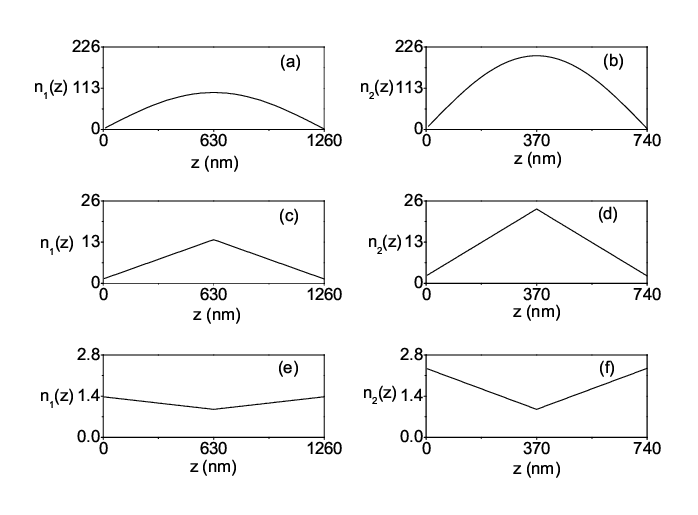}
\caption{The picture of three kinds of functions refractive
indexes in a period. } \label{Fig1}
\end{figure}

\vskip 8pt {\bf 6. Numerical result} \vskip 8pt

We report in this section our numerical results of transmissivity
and dispersion relation. We consider three kinds of functions form
refractive indexes in a period,

(1) The first one is sine type function refractive indexes, as
\begin{eqnarray}
\left \{
\begin{array}{cc}
n_{1}(z)=n_{1}(0)+A_1\sin\frac{\pi}{b}z, \hspace{0.1in} 0 \leq z\leq b\\
n_{2}(z)=n_{2}(0)+A_2\sin\frac{\pi}{a}z, \hspace{0.1in} 0 \leq
z\leq a
\end{array}
\right.
\end{eqnarray}
which are shown in FIG. 4 (a) and (b).

 (2) The second one is upward fold line type function refractive
indexes, as
\begin{eqnarray}
n_{1}(z)= \left \{ \begin{array}{ll}
   n_{1}(0)+\frac{2(m-1)n_1(0)}{b}z, \hspace{0.4in} 0 \leq z\leq \frac{b}{2}, \\
   n_{1}(0)+\frac{2(m-1)n_1(0)}{b}(b-z), \hspace{0.1in} \frac{b}{2} \leq z\leq b,
   \end{array}
   \right.
\end{eqnarray}
 and
\begin{eqnarray}
n_{2}(z)= \left \{ \begin{array}{ll}
   n_{2}(0)+\frac{2(m-1)n_1(0)}{a}z, \hspace{0.4in} 0 \leq z\leq \frac{a}{2}, \\
   n_{2}(0)+\frac{2(m-1)n_1(0)}{a}(a-z), \hspace{0.1in} \frac{a}{2} \leq z\leq a,
   \end{array}
   \right.
\end{eqnarray}
which are shown in FIG. 4 (c) and (d).

 (3) The third one is
downward fold line type function refractive indexes, as
\begin{eqnarray}
n_{1}(z)= \left \{ \begin{array}{ll}
   n_{1}(0)-\frac{2n_1(0)(1-1.1\sin\theta_{t}^{I})}{b}z, \hspace{0.4in} 0 \leq z\leq \frac{b}{2}, \\
   n_{1}(0)-\frac{2n_1(0)(1-1.1\sin\theta_{t}^{I})}{b}(b-z), \hspace{0.1in} \frac{b}{2} \leq z\leq b,
   \end{array}
   \right.
\end{eqnarray}
 and
\begin{eqnarray}
n_{1}(z)= \left \{ \begin{array}{ll}
   n_{1}(0)-\frac{2n_1(0)(1-1.1\sin\theta_{t}^{II})}{a}z, \hspace{0.4in} 0 \leq z\leq \frac{a}{2}, \\
   n_{1}(0)-\frac{2n_1(0)(1-1.1\sin\theta_{t}^{II})}{a}(a-z), \hspace{0.1in} \frac{a}{2} \leq z\leq a,
   \end{array}
   \right.
\end{eqnarray}
which are shown in FIG. 4 (e) and (f).

While $n_{1}(0)$, $n_{2}(0)$, $m$, $A_1$ and $A_2$ are constants,
$b$ and $a$ are half period thickness. With Eq. (84), we can
investigate the dispersion relation and band gap structure, and
can resolve transmissivity from Eqs. (85) and (86). In FIG. 5, we
take sine type function refractive indexes (Eq. (87)), and the
parameters are $\theta_{i}^{0}=\frac{\pi}{3}$, $A_1=100$,
$A_2=200$, $n_1(0)=\sqrt{1.9}$, $n_2(0)=\sqrt{5.5}$, $a=740 nm$
and $b=1260 nm$. The FIG. 5 (a) is the dispersion relation and
FIG. 5 (b) is the transmissivity. In the two figures, we can find
sine function type photonic crystals has band gap structure. In
FIG. 6, we take upward fold line type function refractive indexes
(Eqs. (88) and (89)), and the parameters are:
$\theta_{i}^{0}=\frac{\pi}{3}$, $n_1(0)=\sqrt{1.9}$,
$n_2(0)=\sqrt{5.5}$, $a=740 nm$, $b=1260 nm$ and $m=10$. The FIG.
6 (a) is the dispersion relation and FIG. 6 (b) is the
transmissivity. The two figures show the band gap structure when
refractive index is upward fold line type function. In FIG. 7,
there is band gap structure when the refractive index is taken
downward fold line type (Eqs. (90) and (91)), and the parameters
are $\theta_{i}^{0}=\frac{\pi}{3}$, $n_1(0)=\sqrt{1.9}$,
$n_2(0)=\sqrt{5.5}$, $a=740 nm$,  $b=1260 nm$,
$\sin\theta_{t}^{I}=\frac{n_{0}}{n_{1}(0)}\sin\theta_{i}^{0}$,
$\sin\theta_{t}^{II}=\frac{n_{0}}{n_{2}(0)}\sin\theta_{i}^{0}$ and
$n_{0}=1$. In FIG. 8, we compare the band gap structures of
function photonic crystals with the conventional photonic
crystals. The FIG. 8 (a) is conventional photonic crystals band
gap structures, the parameters are:
$\theta_{i}^{0}=\frac{\pi}{3}$, $n(a)=\sqrt{1.9}$,
$n(b)=\sqrt{5.5}$, $a=740 nm$, and $b=1260 nm$. The FIG. 8 (b) is
sine type function photonic crystals band gap structures, the
parameters are: $\theta_{i}^{0}=\frac{\pi}{3}$, $A_1=0.01$,
$A_2=0.02$, $n_1(0)=\sqrt{1.9}$, $n_2(0)=\sqrt{5.5}$, $a=740 nm$
and $b=1260 nm$. The FIG. 8 (c) is also sine type function
photonic crystals band gap structures, the parameters are:
$\theta_{i}^{0}=\frac{\pi}{3}$, $A_1=100$, $A_2=200$,
$n_1(0)=\sqrt{1.9}$, $n_2(0)=\sqrt{5.5}$, $a=740 nm$ and $b=1260
nm$. The FIG. 8 (d) is upward fold line type function photonic
crystals band gap structure, and the parameters are:
$\theta_{i}^{0}=\frac{\pi}{3}$, $n_1(0)=\sqrt{1.9}$,
$n_2(0))=\sqrt{5.5}$, $a=740 nm$, $b=1260 nm$ and $m=10$. The FIG.
8 (e) is downward fold line type function photonic crystals band
gap structure, and the parameters are:
$\theta_{i}^{0}=\frac{\pi}{3}$, $n_1(0)=\sqrt{1.9}$,
$n_2(0)=\sqrt{5.5}$, $a=740 nm$,  $b=1260 nm$,
$\sin\theta_{t}^{I}=\frac{n_{0}}{n_{1}(0)}\sin\theta_{i}^{0}$,
$\sin\theta_{t}^{II}=\frac{n_{0}}{n_{2}(0)}\sin\theta_{i}^{0}$ and
$n_{0}=1$. In FIG. 8 (c) and (d), the band gap are more wider than
the conventional photonic crystals. In FIG. 8 (b) and (e), the
band gap are more narrower than the conventional photonic
crystals. In order to satisfy different application, we can design
the different kind of function photonic crystals by choosing
different refractive index function form.

\newpage

 \vskip 10pt

{\bf 7. Conclusion} \vskip 8pt In conclusion, we present a new
kind of function photonic crystals, which refractive index is a
function of space position. Unlike conventional PCs, which
structure grow from two materials A and B, with different
dielectric constants $\varepsilon_{A}$ and $\varepsilon_{B}$.
Based on Fermat principle, we achieve the motion equations of
light in one-dimensional, two-dimensional and three-dimensional
function photonic crystals. For one-dimensional function photonic
crystals, we investigate the dispersion relation, band gap
structure and transmissivity, and compare them with conventional
photonic crystals. By choosing different refractive index
distribution function $n(z)$, we can obtain more wider or more
narrower band gap structure than conventional photonic crystals.
Due to the function photonic crystals has more wider or more
narrower band gap structure than conventional photonic crystals,
we think the function photonic crystals should has more extensive application foreground.\\

\newpage

\begin{figure}[tbp]
\includegraphics[width=10 cm]{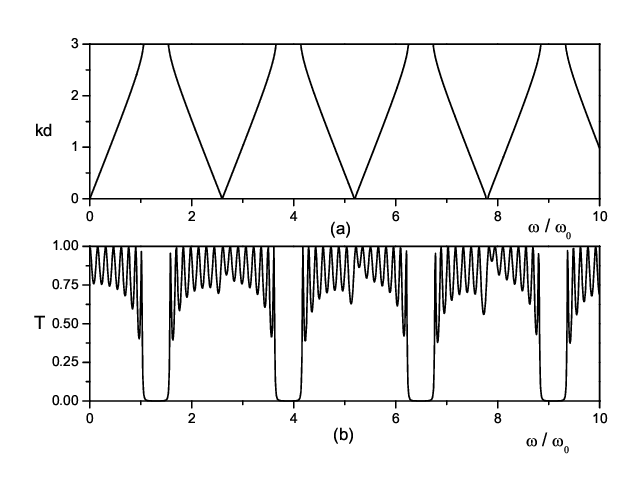}
\caption{The dispersion relation, band gap structure and
transmissivity for sine type function refractive indexes (Eq.
(87)). } \label{Fig1}
\end{figure}

\begin{figure}[tbp]
\includegraphics[width=10 cm]{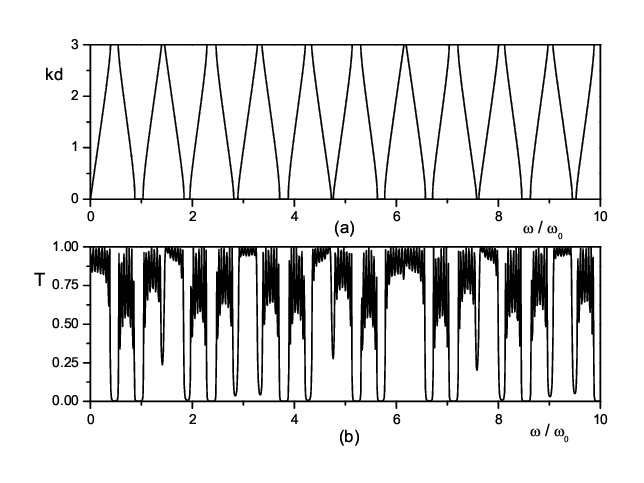}
\caption{The dispersion relation, band gap structure and
transmissivity for upward fold line function refractive indexes
(Eqs. (88)and (89)).} \label{Fig1}
\end{figure}

\begin{figure}[tbp]
\includegraphics[width=10 cm]{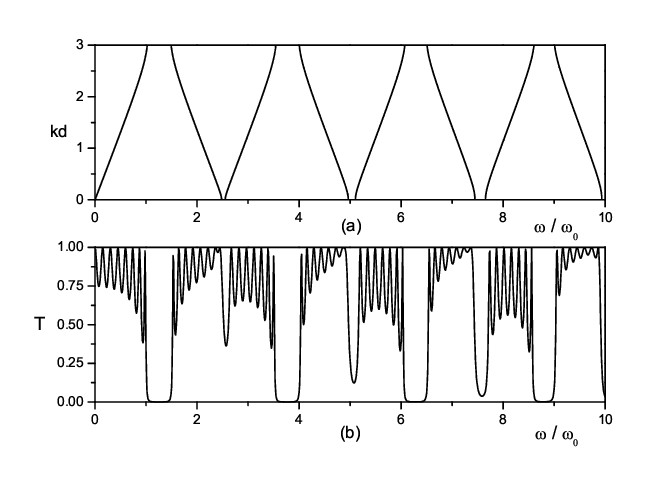}
\caption{The dispersion relation, band gap structure and
transmissivity for downward fold line function refractive indexes
(Eqs. (90)and (91)).} \label{Fig1}
\end{figure}

\begin{figure}[tbp]
\includegraphics[width=10 cm]{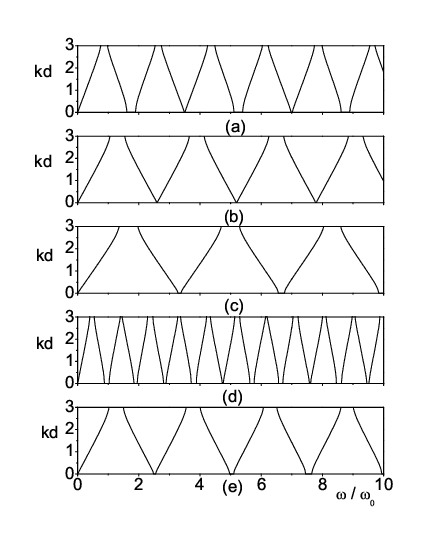}
\caption{Compare the band gap structures of different refractive
index distribution function $n(z)$ function photonic\\
crystals with conventional photonic crystals.} \label{Fig1}
\end{figure}

\end{document}